\documentclass[12pt]{article}
\usepackage{amssymb}

\input{tcilatex}
\begin{document}

\bigskip\ 

\bigskip\ 

\begin{center}
\textbf{QUBITS AND CHIROTOPES}

\textbf{\ }

\textbf{\ }

\smallskip\ 

J. A. Nieto\footnote{%
nieto@uas.uasnet.mx, janieto1@asu.edu}

\smallskip

\textit{Mathematical, Computational \& Modeling Science Center, Arizona
State University, PO Box 871904, Tempe, AZ 85287, USA}

\textit{Facultad de Ciencias F\'{\i}sico-Matem\'{a}ticas de la Universidad
Aut\'{o}noma de Sinaloa, 80010, Culiac\'{a}n Sinaloa, M\'{e}xico}

\textit{Departamento de Investigaci\'{o}n en F\'{\i}sica de la Universidad
de Sonora, 83000, Hermosillo Sonora , M\'{e}xico}

\bigskip\ 

\bigskip\ 

Abstract
\end{center}

We show that qubit and chirotope concepts are closely related. In fact, we
prove that the qubit concept leads to a generalization of the chirotope
concept, which we call qubitope. Moreover, we argue that a possible qubitope
theory may suggest interesting applications of oriented matroid theory in at
least three physical contexts, in which qubits make their appearance, namely
string theory, black holes and quantum information.

\bigskip\ 

\bigskip\ 

\bigskip\ 

\bigskip\ 

\bigskip\ 


Pacs numbers: 04.60.-m, 04.65.+e, 11.15.-q, 11.30.Ly

June, 2010

\newpage

Recently, in a number of remarkable developments [1]-[8], a relation
between, apparently two different scenarios, black holes and quantum
information, has been established. The key concept for this link has been
the so called quantum bit notion, or qubit, which is the smallest unit of
quantum information. In appropriate qubit basis, the components of a pure
state $\mid \psi >$ can be written in terms hypermatrix $%
a_{a_{1}a_{2}...a_{N}}$ which in turn leads to a density matrix $\rho $. It
turns out that $\rho $ can be defined in terms of the hyperdeterminant
associated with $a_{a_{1}a_{2}...a_{N}}$, a quantity introduced for the
first time by Cayley in 1845 [9]. Surprisingly, in some cases the quantity $%
a_{a_{1}a_{2}...a_{N}}$ can also be related to the  entropy of STU black
holes via also its hyperdeterminant (see Ref. [4] for details).

On the other hand, it is known that the chirotope concept plays a
fundamental role in oriented matroid theory [10]. In fact, the emergence of
this concept can be traced back to the origin of matroids [11] which can be
understood as a generalization of matrices. From a modern perspective,
however, one may introduce the mathematical notion of chirotope, or oriented
matroid, by considering a generalization of the Grassmann-Pl\"{u}cker
relations of ordinary determinants [12].

Thus, we have two generalizations of the matrix notion, namely hypermatrix
and matroid. Since a qubit is related with hyperdeterminants of
hypermatrices and a chirotope is connected with a generalization of ordinary
determinants via the Grassmann-Pl\"{u}cker relations one may wonder whether
these two qubit-chirotope concepts are related. If we achieve such a
relation then one may be in a position to bring a variety of mathematical
tools from oriented matroid theory to black-hole physics and vice versa.

Our starting point is to consider a possible scenario in which the qubit
concept makes its appearance [1], namely the (2+2)-signature flat target
\textquotedblleft spacetime\textquotedblright\ of the Nambu-Goto action. Let
us first observe that the line element,%
\begin{equation}
ds^{2}=dx^{\mu }dx^{\nu }\eta _{\mu \nu },  \tag{1}
\end{equation}%
of flat space with (2+2)-signature, with $\eta _{\mu \nu }=diag(-1,-1,1,1)$,
may also be written as

\begin{equation}
ds^{2}=\frac{1}{2}dx^{ab}dx^{cd}\varepsilon _{ac}\varepsilon _{bd},  \tag{2}
\end{equation}%
where the matrix coordinates $x^{ab}$ are given by

\begin{equation}
x^{ab}=\left( 
\begin{array}{cc}
x^{1}+x^{3} & x^{2}+x^{4} \\ 
x^{2}-x^{4} & -x^{1}+x^{3}%
\end{array}%
\right) ,  \tag{3}
\end{equation}%
and $\varepsilon _{ab}$ is the completely antisymmetric symbol with $%
\varepsilon _{12}=1$.

Similarly, it is not difficult to show [1] (see also Ref. [2]) that the
world sheet metric

\begin{equation}
\gamma _{ab}=\partial _{a}x^{\mu }\partial _{b}x^{\nu }\eta _{\mu \nu }, 
\tag{4}
\end{equation}%
can also be written as

\begin{equation}
\gamma _{ab}=\frac{1}{2}\partial _{a}x^{cd}\partial _{b}x^{ef}\varepsilon
_{ce}\varepsilon _{df}.  \tag{5}
\end{equation}%
This expression motivates to write the determinant of $\gamma _{AB}$,

\begin{equation}
\det \gamma =\frac{1}{2}\varepsilon ^{ab}\varepsilon ^{cd}\gamma _{ac}\gamma
_{bd},  \tag{6}
\end{equation}%
in the form

\begin{equation}
\det \gamma =\frac{1}{2}\varepsilon ^{ab}\varepsilon ^{cd}\varepsilon
_{eg}\varepsilon _{fh}\varepsilon _{ru}\varepsilon
_{sv}a_{a}^{ef}a_{c}^{gh}a_{b}^{rs}a_{d}^{uv}=Deta,  \tag{7}
\end{equation}%
with

\begin{equation}
a_{a}^{cd}\equiv \partial _{a}x^{cd}.  \tag{8}
\end{equation}%
One recognizes in (7) the hyperdeterminant of the hypermatrix $a_{a}^{cd}$.
Thus, this proves that the Nambu-Goto action [13]-[14]

\begin{equation}
S=\frac{1}{2}\int d^{2}\xi \sqrt{\det \gamma },  \tag{9}
\end{equation}%
for a flat target \textquotedblleft spacetime\textquotedblright\ with
(2+2)-signature can also be written as [1]

\begin{equation}
S=\frac{1}{2}\int d^{2}\xi \sqrt{Deta}.  \tag{10}
\end{equation}

We shall now show that the hyperdeterminant (7) can be linked to the
chirotope concept. For this purpose by using (4) we first write (6) in the
alternative Schild type [15] form

\begin{equation}
\det \gamma =\frac{1}{2}\sigma ^{\mu \nu }\sigma ^{\alpha \beta }\eta _{\mu
\alpha }\eta _{\nu \beta },  \tag{11}
\end{equation}%
where

\begin{equation}
\sigma ^{\mu \nu }=\varepsilon ^{ab}a_{a}^{\mu }a_{b}^{\nu }.  \tag{12}
\end{equation}%
Here, we have used the definition

\begin{equation}
a_{a}^{\mu }=\partial _{a}x^{\mu }.  \tag{13}
\end{equation}%
It turns out that the quantity $\chi ^{\mu \nu }=sign\sigma ^{\mu \nu }$ is
a chirotope of an oriented matroid (see Refs. [16-18]). In fact, since $%
\sigma ^{\mu \nu }$ satisfies the identity

\begin{equation}
\sigma ^{\mu \lbrack \nu }\sigma ^{\alpha \beta ]}\equiv 0,  \tag{14}
\end{equation}%
one can verify that $\chi ^{\mu \nu }$ satisfies the Grassmann-Pl\"{u}cker
relation

\begin{equation}
\chi ^{\mu \lbrack \nu }\chi ^{\alpha \beta ]}=0,  \tag{15}
\end{equation}%
and therefore $\chi ^{\mu \nu }$ is a realizable chirotope (see Ref. [10]
and references therein). Here, the bracket $[\nu \alpha \beta ]$ means
completely antisymmetric.

Since the Grassmann-Pl\"{u}cker relation (15) holds, the ground set 
\begin{equation}
E=\{\mathbf{1,2,3,4}\}  \tag{16}
\end{equation}%
and the alternating map 
\begin{equation}
\chi ^{\mu \nu }\rightarrow \{-1,0,1\},  \tag{17}
\end{equation}%
determine a $2$-rank realizable oriented matroid $M=(E,\chi ^{\mu \nu })$.
The collection of bases for this oriented matroid is

\begin{equation}
\mathcal{B}=\{\mathbf{\{1,2\},\{1,3\},\{1,4\},\{2,3\},\{2,4\},\{3,4\}}\}, 
\tag{18}
\end{equation}%
which can be obtained by just given values to the indices $\mu $ and $\nu $
in $\chi ^{\mu \nu }$. Actually, the pair $(E,\mathcal{B})$ determines a $2$%
-rank uniform nonoriented ordinary matroid.

Using the definition

\begin{equation}
\sigma ^{efrs}\equiv \varepsilon ^{ab}a_{a}^{ef}a_{b}^{rs},  \tag{19}
\end{equation}%
one can show that the hyperdeterminant (7) can also be written as

\begin{equation}
\det \gamma =\frac{1}{2}\sigma ^{efrs}\sigma ^{ghuv}\varepsilon
_{eg}\varepsilon _{fh}\varepsilon _{ru}\varepsilon _{sv}=Deta.  \tag{20}
\end{equation}%
So, we have achieved our goal of writing the hyperdeterminant (7) in terms
of a \textquotedblleft chirotope\textquotedblright\ structure (19). Our
strategy was to translate the \textquotedblleft chirotope\textquotedblright\
given in (12) to the form (19). However, by comparing (12) and (19) one
finds that there are important differences between these two expressions
which suggest a possible generalization of the chirotope concept. From (12)
we obtain the property%
\begin{equation}
\sigma ^{\mu \nu }=-\sigma ^{\nu \mu },  \tag{21}
\end{equation}%
that is, $\sigma ^{\mu \nu }$ is completely antisymmetric (alternative)
quantity, while in (19) we have the weaker condition 
\begin{equation}
\sigma ^{efrs}=-\sigma ^{rsef}.  \tag{22}
\end{equation}%
This means that that the quantity $\sigma ^{efrs}$, which we shall call
qubitope (qubit-chirotope), is not completely antisymmetric but only
alternative in pair of indices. Further, while in the case of (12) the
ground set $E$ is given by (16), the expressions (19) and (20) suggests to
introduce the underlying ground bitset (from bit and set)%
\begin{equation}
\mathcal{E}=\{1,2\}  \tag{23}
\end{equation}%
and the pre-ground set

\begin{equation}
E_{0}=\{(1,1),(1,2),(2,1),(2,2)\}.  \tag{24}
\end{equation}%
So, our task is to find the relation between $E_{0}$ and $E$. By comparing
(16) and (24) one sees that by establishing the labels

\begin{equation}
\begin{array}{cc}
(1,1)\leftrightarrow \mathbf{1}, & (1,2)\leftrightarrow \mathbf{2}, \\ 
&  \\ 
(2,1)\leftrightarrow \mathbf{3}, & (2,2)\leftrightarrow \mathbf{4}.%
\end{array}
\tag{25}
\end{equation}%
such a relation is achieved. This can be understood considering that (25) is
equivalent to make the identification of indices $\{a,b\}\leftrightarrow \mu 
$,..,etc. Observe that considering this identifications the family of bases
(18) becomes

\begin{equation}
\begin{array}{c}
\mathcal{B}_{0}=\{\{(1,1),(1,2)\},\{(1,1),(2,1)\},\{(1,1),(2,2)\}, \\ 
\\ 
\{(1,2),(2,1)\},\{(1,2),(2,2)\},\{(2,1),(2,2)\}\}.%
\end{array}
\tag{26}
\end{equation}%
Thus, from the qubitope $\sigma ^{efrs}$, we have discovered the underlying
structure $Q=(\mathcal{E},E_{0},B_{0})$. By convenience we call this new
structure $Q$ a qubitoid. The word \textquotedblleft
qubitoid\textquotedblright\ is short for qubit-matroid.

Let us try to generalize the above scenario to higher dimensions. First, we
would like to extend the steps in the expressions (1) and (2). If we
consider the coordinates $x^{abc}$ instead of $x^{ab}$ one finds that the
null line element

\begin{equation}
ds^{2}=\frac{1}{2}dx^{abc}dx^{def}\varepsilon _{ad}\varepsilon
_{be}\varepsilon _{cf},  \tag{27}
\end{equation}%
vanishes identically. This follows because $dx^{abc}dx^{def}\varepsilon
_{ad}\varepsilon _{be}=s^{cf}$ is a symmetric quantity, while $\varepsilon
_{cf}$ is antisymmetric. Similarly one can verify that the hyperdeterminant
of the hypermatrix $a_{a}^{bcd}\equiv \partial _{a}x^{bcd}$ present some
difficulties due to the fact that the analogue of (7) can not be obtained.
In fact, the quantity $\lambda _{ab}=\partial _{a}x^{efg}\partial
_{b}x^{hrs}\varepsilon _{eh}\varepsilon _{fr}\varepsilon _{gs}$ is
antisymmetric rather than symmetric as the metric $\gamma _{ab}$ and
therefore in this case the steps (4) and (5) can not follow. Hence, from the
Nambu-Goto action point of view this case, which corresponds to
(4+4)-signature, is not very interesting, although in the Polyakov action
context may still be interesting. So, we jump to the next possibility,
namely the line element

\begin{equation}
ds^{2}=\frac{1}{2}dx^{abcr}dx^{defs}\varepsilon _{ad}\varepsilon
_{be}\varepsilon _{cf}\varepsilon _{rs},  \tag{28}
\end{equation}%
which, one can show, implies a line element of the type (1), but now
associated with a flat target (8+8)-signature \textquotedblleft
spacetime\textquotedblright . Explicitly, we have the relations%
\begin{equation}
\begin{array}{ccc}
x^{1111}\leftrightarrow x^{1}+x^{9}, & x^{2222}\leftrightarrow -x^{1}+x^{9},
& x^{1112}\leftrightarrow x^{2}+x^{10}, \\ 
&  &  \\ 
x^{2221}\leftrightarrow x^{2}-x^{10}, & x^{1122}\leftrightarrow x^{3}+x^{11},
& x^{2211}\leftrightarrow -x^{3}+x^{11}, \\ 
&  &  \\ 
x^{1121}\leftrightarrow x^{4}+x^{12}, & x^{2212}\leftrightarrow x^{4}-x^{12},
& x^{1212}\leftrightarrow x^{5}+x^{13}, \\ 
&  &  \\ 
x^{2121}\leftrightarrow -x^{5}+x^{13}, & x^{1211}\leftrightarrow
x^{6}+x^{14}, & x^{2122}\leftrightarrow x^{6}-x^{14}, \\ 
&  &  \\ 
x^{1221}\leftrightarrow x^{7}+x^{15}, & x^{2112}\leftrightarrow
-x^{7}+x^{15}, & x^{1222}\leftrightarrow x^{8}+x^{16}, \\ 
&  &  \\ 
x^{2111}\leftrightarrow x^{8}-x^{16}. &  & 
\end{array}
\tag{29}
\end{equation}%
In this case, the hyperdeterminant of the hypermatrix%
\begin{equation}
a_{a}^{bcde}=\partial _{a}x^{bcde}  \tag{30}
\end{equation}%
is given by (see Eq. (32) of Ref. [19])

\begin{equation}
\begin{array}{c}
\det \gamma = \\ 
\\ 
\frac{1}{2}\varepsilon ^{ab}\varepsilon ^{cd}\varepsilon
_{e_{1}f_{1}}\varepsilon _{e_{2}f_{2}}\varepsilon _{e_{3}f_{3}}\varepsilon
_{e_{4}f_{4}}\varepsilon _{g_{1}h_{1}}\varepsilon _{g_{2}h_{2}}\varepsilon
_{g_{3}h_{3}}\varepsilon
_{g_{4}h_{4}}a_{a}^{e_{1}e_{2}e_{3}e_{4}}a_{c}^{f_{1}f_{2}f_{5}f_{4}}a_{b}^{g_{1}g_{2}g_{3}g_{4}}a_{d}^{h_{1}h_{2}h_{3}h_{4}}
\\ 
\\ 
=Deta.%
\end{array}
\tag{31}
\end{equation}%
Thus, by substituting (31) into (10) we find a Nambu-Goto action for a flat
target \textquotedblleft spacetime\textquotedblright\ with (8+8)-signature
written in terms of the hyperdeterminant $Deta$.

The qubitoid now is determined by the underlying set

\begin{equation}
\mathcal{E}=\{1,2\},  \tag{32}
\end{equation}%
and the pre-ground set

\begin{equation}
\begin{array}{cc}
E_{0}= & \{(1,1,1,1),(1,1,1,2),(1,1,2,1),(1,1,2,2), \\ 
&  \\ 
& (1,2,1,1),(1,2,1,2),(1,2,2,1),(1,2,2,2) \\ 
&  \\ 
& (2,1,1,1),(2,1,1,2),(2,1,2,1),(2,1,2,2) \\ 
&  \\ 
& (2,2,1,1),(2,2,1,2),(2,2,2,1),(2,2,2,2)\}.%
\end{array}
\tag{33}
\end{equation}%
It is not difficult to see that by making the identifications

\begin{equation}
\begin{array}{ccc}
(1,1,1,1)\leftrightarrow \mathbf{1} & (1,1,1,2)\leftrightarrow \mathbf{2} & 
(1,1,2,1)\leftrightarrow \mathbf{3} \\ 
&  &  \\ 
(1,1,2,2)\leftrightarrow \mathbf{4} & (1,2,1,1)\leftrightarrow \mathbf{5} & 
(1,2,1,2)\leftrightarrow \mathbf{6} \\ 
&  &  \\ 
(1,2,2,1)\leftrightarrow \mathbf{7} & (1,2,2,2)\leftrightarrow \mathbf{8} & 
(2,1,1,1)\leftrightarrow \mathbf{9} \\ 
&  &  \\ 
(2,1,1,2)\leftrightarrow \mathbf{10} & (2,1,2,1)\leftrightarrow \mathbf{11}
& (2,1,2,2)\leftrightarrow \mathbf{12} \\ 
&  &  \\ 
(2,2,1,1)\leftrightarrow \mathbf{13} & (2,2,1,2)\leftrightarrow \mathbf{14}
& (2,2,2,1)\leftrightarrow \mathbf{15} \\ 
&  &  \\ 
(2,2,2,2)\leftrightarrow \mathbf{16,} &  & 
\end{array}
\tag{34}
\end{equation}%
one obtains a relation between the pre-ground set $E_{0}$ given in (33) and
the ground set 
\begin{equation}
E=\{\mathbf{1,2,...,15,16}\}.  \tag{35}
\end{equation}%
This can be understood by considering that (34) is equivalent to make the
identification of indices $(a,b,c,d)\leftrightarrow \mu ,...etc$. It turns
out that considering these relations one finds that the collection of bases $%
\mathcal{B}$ contains $\left( 
\begin{array}{c}
16 \\ 
2%
\end{array}%
\right) =120$ two-element subset of the 16-element set $E$, given in (35).
This $2$-element subset can be obtained by considering a lexicographic order
of all 120 two-subsets of $\{\mathbf{1,2,...,15,16}\}$. For instance, the
first 35 two-subsets of $\mathcal{B}$ are 
\begin{equation}
\begin{array}{ccccccc}
\mathbf{\{1,2\},} & \mathbf{\{1,3\},} & \mathbf{\{1,4\},} & \mathbf{\{1,5\},}
& \mathbf{\{1,6\},} & \mathbf{\{1,7\},} & \mathbf{\{1,8\},} \\ 
\mathbf{\{1,9\},} & \mathbf{\{1,10\},} & \mathbf{\{1,11\},} & \mathbf{%
\{1,12\},} & \mathbf{\{1,13\},} & \mathbf{\{1,14\},} & \mathbf{\{1,15\},} \\ 
\mathbf{\{1,16\},} & \mathbf{\{2,3\},} & \mathbf{\{2,4\},} & \mathbf{\{2,5\},%
} & \mathbf{\{2,6\},} & \mathbf{\{2,7\},} & \mathbf{\{2,8\},} \\ 
\mathbf{\{2,9\},} & \mathbf{\{2,10\},} & \mathbf{\{2,11\},} & \mathbf{%
\{2,12\},} & \mathbf{\{2,13\},} & \mathbf{\{2,14\},} & \mathbf{\{2,15\},} \\ 
\mathbf{\{2,16\},} & \mathbf{\{3,4\},} & \mathbf{\{3,5\},} & \mathbf{\{3,6\},%
} & \mathbf{\{3,7\},} & \mathbf{\{3,8\},} & \mathbf{\{3,9\},...}%
\end{array}
\tag{36}
\end{equation}%
The sequence follows until the last term $\mathbf{\{15,16\}}$. By using (34)
one finds that the first terms of $\mathcal{B}_{0}$ look like

\begin{equation}
\begin{array}{c}
\mathcal{B}_{0}=\{\{(1,1,1,1),\{1,1,1,2)\},\{(1,1,1,1),\{1,1,2,1)\}, \\ 
\\ 
\{(1,1,1,1),\{1,1,2,2)\},\{(1,1,1,1),\{1,2,1,1)\},\{(1,1,1,1),\{1,2,1,2)%
\},...\}.%
\end{array}
\tag{37}
\end{equation}%
Thus, associated with the the quantity $a_{a}^{bcde}$ we have again a
qubitoid structure of the form $Q=(\mathcal{E},E_{0},B_{0})$ which
corresponds to a flat target \textquotedblleft spacetime\textquotedblright\
of (8+8)-dimensions. The corresponding qubitope is given by%
\begin{equation}
\sigma ^{a_{1}a_{2}a_{3}a_{4}a_{5}a_{6}a_{7}a_{8}}=\frac{1}{2}\varepsilon
^{bc}a_{b}^{a_{1}a_{2}a_{3}a_{4}}a_{c}^{a_{5}a_{6}a_{7}a_{8}}.  \tag{38}
\end{equation}%
It is worth mentioning that while in (2+2)-dimensions the quantity $Deta$ is
invariant under $SL(2,R)^{3}$ in the case of (8+8)-dimensions $Deta$ must be
invariant under $SL(2,R)^{5}$.

The method, of course, can be extended to $(\frac{2^{2n+2}}{2}+\frac{2^{2n+2}%
}{2})$-signature, $n=0,1,2,...etc$. For the cases of $(\frac{2^{2n+1}}{2}+%
\frac{2^{2n+1}}{2})$-signature the corresponding line element vanishes
identically.

It remains to explore whether the present qubitoid and qubitope formalism
will allow us a deeper understanding of other two scenarios, namely
black-holes and quantum information. In the first case, as in
(2+2)-dimensions, one may think in a black hole with $\frac{2^{2n+2}}{2}$%
-electric and $\frac{2^{2n+2}}{2}$-magnetic charges with entropy

\begin{equation}
S=\pi \sqrt{Deta}.  \tag{39}
\end{equation}%
While in the second case, one may introduce pure states $\mid \psi >$
associated with the $\frac{2^{2n+2}}{2}$-qubitoid system. For instance in
the case of (8+8)-dimensions the pure states $\mid \psi >$ must be given by
(see Refs [19] and [20])

\begin{equation}
\mid \psi
>=\dsum\limits_{a_{1}a_{2}a_{3}a_{4}a_{5}}a_{a_{1}}^{a_{2}a_{3}a_{4}a_{5}}%
\mid a_{1}a_{2}a_{3}a_{4}a_{5}>.  \tag{40}
\end{equation}

It is worth mentioning that the complete classification of $N$-qubit systems
is a difficult, or perhaps an impossible, task. In reference [19] an
interesting development for characterizing a subclass of $N$-qubit
entanglement has been considered. An attractive aspect of this construction
is that the $N$-qubit entanglement can be understood in geometric terms. The
idea is based on the bipartite partitions of the Hilbert space in the form $%
C^{2^{N}}=C^{L}\otimes C^{l}$, with $L=2^{N-n}$ and $l=2^{n}$. Such a
partition allows a geometric interpretation in terms of the complex
Grassmannian variety $Gr(L,l)$ of $l$-planes in $C^{L}$ via the Pl\"{u}cker
embedding. In this case, the Plucker coordinates of the Grassmannians are
natural invariants of the theory. In this scenario the $5$-qubit given in
(40) admit a geometric interpretation in terms of the complex Grassmannian $%
Gr(8,4)$. Considering such an interpretation it has been proved that the
expression (31) is an hyperdeterminant associated with the Plucker
coordinates of the Grassmannian $Gr(8,4)$ (see eq. (32) of Ref. [19] and
Ref. [20] for $5$-qubit discussion).

Furthermore, it is interesting that the line element (28) also appears on
several physical contexts. First of all, extremal black hole solutions in
the STU model of $D=4$, $\mathcal{N}=2$ supergravity admit a description in
terms of $4$-qubit systems [21]-[22] (for a $4$-qubit entanglement see [23]
and references therein). In this case, the line element corresponds to the
moduli space $\mathcal{M}_{4}=[U(1)\backslash SL(2;R)]^{3}$. rather than to
the \textquotedblleft spacetime\textquotedblright . Upon dimensional
reduction $\mathcal{M}_{4}$ becomes $\mathcal{M}_{3}=[SO(4)]^{2}\backslash
SO(4,4)$ or $\mathcal{M}_{3}^{\ast }=[SO(2,2)]^{2}\backslash SO(4,4)$
depending whether the truncation is along a space-like or time-like
direction, respectively. Among other things, the relevance of this
construction in our approach is that the signature of the metric $\mathcal{M}%
_{3}^{\ast }$ is also of the type (8+8) (see Refs. [21] and [22] for
details).

It is remarkable that the Nambu-Goto action in a flat target
\textquotedblleft spacetime\textquotedblright\ with $(\frac{2^{2n+2}}{2}+%
\frac{2^{2n+2}}{2})$-signature emerges as the underlying motivation for
studying the new mathematical structures of qubitoids $Q=(\mathcal{E}%
,E_{0},B_{0})$ and the corresponding qubitopes.

\bigskip\ 

\noindent \textbf{Acknowledgments: }I would like to thank M. C. Mar\'{\i}n
and A. Le\'{o}n for helpful comments and the Mathematical, Computational \&
Modeling Science Center of the Arizona State University where part of this
work was developed.

\smallskip\


\begin{thebibliography}{99}
\bibitem{1} M. J. Duff, Phys. Lett. B \textbf{641}, 335 (2006); arXiv:
hep-th/0602160.

\bibitem{2} J. A. Nieto, Mod. Phys. Lett. A \textbf{22}, 2453 (2007); arXiv:
hep-th/0606219.

\bibitem{3} M. J. Duff and S. Ferrara, Lect. Notes \ Phys. \textbf{755}, 93
(2008); arXiv: hep-th/0612036.

\bibitem{4} L. Borsten, D. Dahanayake, M. J. Duff, H. Ebrahim, W. Rubens,
Phys. Rept. \textbf{471}, 113 (2009); arXiv: hep-th/0809.4685.

\bibitem{5} H. Nishino and S. Rajpoot, Phys. Lett. B \textbf{652}, 135
(2007); arXiv: 0709.0973.

\bibitem{6} L. Castellani, P. A. Grassi and L. Sommovigo, Phys. Lett. B 
\textbf{678}, 308 (2009); arXiv: 0904.2512.

\bibitem{7} R. Kallosh and A. D. Linde, Phys. Rev. D \textbf{73}, 104033
(2006); arXiv: hep-th/0602061.

\bibitem{8} P. L\'{e}vay, Phys. Rev. D \textbf{74}, 024030 (2006); arXiv:
hep-th/0603136.

\bibitem{9} A. Cayley, Camb. Math. J. \textbf{4,} 193 (1845).

\bibitem{10} A. Bj\"{o}rner, M. Las Vergnas, B. Sturmfels, N. White and G.
M. Ziegler, \textit{Oriented Matroids}, (Cambridge University Press,
Cambridge, 1993).

\bibitem{11} H. Whitney, Amer. J. Math. \textbf{57}, 509 (1935).

\bibitem{12} J. P. S., Kung, \textquotedblleft Basis Exchange
Properties\textquotedblright\ in N. L White, ed., (Cambridge University
Press., Cambridge, 1986), pp. 62-75.

\bibitem{13} Y. Nambu, \textquotedblleft Duality and
hydrodynamics,\textquotedblright\ Lectures at the Copenhagen conference,
1970.

\bibitem{14} T. Goto, Prog. Theor. Phys. \textbf{46}, 1560 (1971).

\bibitem{15} A. Schild, Phys. Rev. D \textbf{16}, 1722 (1977).

\bibitem{16} J. A. Nieto, Adv. Theor. Math. Phys. \textbf{8}, 177 (2004);
arXiv: hep-th/0310071.

\bibitem{17} J. A. Nieto, Adv. Theor. Math. Phys. \textbf{10}, 747 (2006),
arXiv: hep-th/0506106.

\bibitem{18} J. A. Nieto, J. Math. Phys. \textbf{45}, 285 (2004); arXiv:
hep-th/0212100.

\bibitem{19} P. Levay, J. Phys. A \textbf{38}, 9075 (2005).

\bibitem{20} J. G. Luque and J. Y. Thibon, \textquotedblleft Algebraic
invariants of five qubits\textquotedblright , arXiv: quant-ph/0506058.

\bibitem{21} G. Bossard, Y. Michel and B. Pioline, JHEP \textbf{1001}, 038
(2010); arXiv: 0908.1742.

\bibitem{22} P. Levay, "STU Black Holes as Four Qubit Systems", arXiv:
1004.3639.

\bibitem{23} J. G. Luque and J. Y. Thibon, Phys. Rev A. \textbf{67}, 042303
(2003); arXiv: quant-ph/0212069.
\end{thebibliography}
\end{document}